\documentclass[pdflatex,sn-nature]{sn-jnl}

\usepackage{graphicx}%
\usepackage{multirow}%
\usepackage{amsmath,amssymb,amsfonts}%
\usepackage{amsthm}%
\usepackage{mathrsfs}%
\usepackage[title]{appendix}%
\usepackage{xcolor}%
\usepackage{textcomp}%
\usepackage{manyfoot}%
\usepackage{booktabs}%
\usepackage[T1]{fontenc}
\usepackage{algorithm}%
\usepackage{algorithmicx}%
\usepackage{algpseudocode}%
\usepackage{listings}%

\theoremstyle{thmstyleone}%
\theoremstyle{thmstyletwo}%

\theoremstyle{thmstylethree}%

\begin{document}
\title[Molecular Pawns on a Surface Chessboard]{Molecular Pawns on a
Surface Chessboard: Computational Investigation of Nanographenes with
Atomistic Methods}

\author[1]{\fnm{Antonio Lorenzo} \sur{Mariano}}

\author[1]{\fnm{Jacopo} \sur{Lupi}}

\author[1]{\fnm{Emilio} \sur{Lorini}}

\author*[1]{\fnm{Matteo} \sur{Briganti}}\email{matteo.briganti@unifi.it}

\affil[1]{\orgdiv{Dipartimento di Chimica ``Ugo Schiff”}, \orgname{Università degli Studi di Firenze},
\orgaddress{\street{Via della Lastruccia 3-13}, \city{Sesto Fiorentino}, \postcode{50019}, \country{Italy}}}

\abstract{Atomically precise nanographenes combine tunable electronic topology with localized, coherent spin states, making them candidate platforms for molecular quantum technologies. This review examines the multiscale theoretical framework governing these systems: the hierarchy of computational methods for describing magnetic properties, the substrate–molecule interface, on-surface synthetic kinetics, and supramolecular self-assembly. We highlight how density functional theory, automated reaction-path discovery, and spin-aware machine-learned interatomic potentials connect static electronic structure to mesoscale dynamics, and how magnetic exchange, optical excitations, and spin transport can be predicted before synthesis.}

\keywords{nanographenes, on-surface synthesis, magnetic exchange coupling,
machine-learned interatomic potentials, reaction networks, self-assembly}

\maketitle

\section{The Categorical Imperative: Atomically Precise Quantum Architectures}\label{sec:imperative}

The realization of scalable and robust quantum computing architectures fundamentally hinges on the precise manipulation of quantum phenomena, most notably superposition and quantum entanglement. While traditional solid-state platforms, such as superconducting circuits and semiconductor quantum dots, have driven the first wave of quantum advancements, they are inherently limited by macroscopic inhomogeneities, substrate noise, and structural defects that rapidly accelerate quantum decoherence \cite{yang2019achieving,burkard2023semiconductor,terhal2015quantum,wehner2018quantum,arute2019quantum}. Top-down fabrication, the standard subtractive manufacturing approach used in the semiconductor industry (such as photolithography and etching), is encountering severe limitations as it attempts to scale quantum computers from dozens to the millions of interacting, high-fidelity qubits required for fault-tolerant quantum computing. To circumvent these bottlenecks and build truly atomically precise quantum logic gates, the field is undergoing a paradigm shift toward molecular systems, where the physical architecture is deterministically controlled at the atomic level through bottom-up chemical synthesis.

Molecular systems \cite{gaita2019molecular,atzori2019second,greule2026exchange}, specifically, atomically precise nanographenes synthesized via bottom-up, on-surface methodologies, offer an ideal and structurally defined platform for this quantum imperative. Unlike top-down etched nanostructures, where defects act as decoherence pathways, bottom-up synthesized nanographenes can achieve a very high degree of structural homogeneity and tunability \cite{mishra2020topological,Mishra2021Nature,vaxevani2026parity}. Furthermore, because carbon lattices possess intrinsically weak spin-orbit coupling (SOC) and can be isotopically purified to eliminate nuclear spin (using pure $^{12}$C), they provide an exceptionally ``quite'' magnetic environment, enabling, in theory, extraordinarily long electron spin coherence times ($T_2$) well above hundreds of microseconds \cite{balasubramanian2009ultralong}.

Crucially, this strict atomic precision is not merely a synthetic achievement; it is a mandatory prerequisite for the control of quantum entanglement \cite{heinrich2021quantum,madzik2022precision,wang2023atomic}. In carbon-based quantum logic gates, information could be encoded in localized molecular orbitals on the edges of the molecular structure (e.g., topological zero-modes on zigzag peripheries). The magnetic exchange interaction ($J$) that dictates the entanglement between these localized spin centers is exponentially sensitive to the spatial distance and the exact conjugated bonding topology connecting them. Even small defects can indeed quench the desired entangled state. 

Therefore, the deterministic engineering of these molecular quantum architectures relies entirely on predictive, high-accuracy computational chemistry and many-body physics. Atomistic modeling serves as an essential tool for computing exchange couplings, mapping spin coupling topologies, and rigorously designing the quantum phase of these highly correlated nanostructures long before their physical realization in an ultra-high-vacuum environment.

\section{Theoretical Foundations of Nanographene Magnetism and Electronic Structure}
\label{sec:theofound}
\begin{figure}[ht]
    \centering
    \includegraphics[width=1\linewidth]{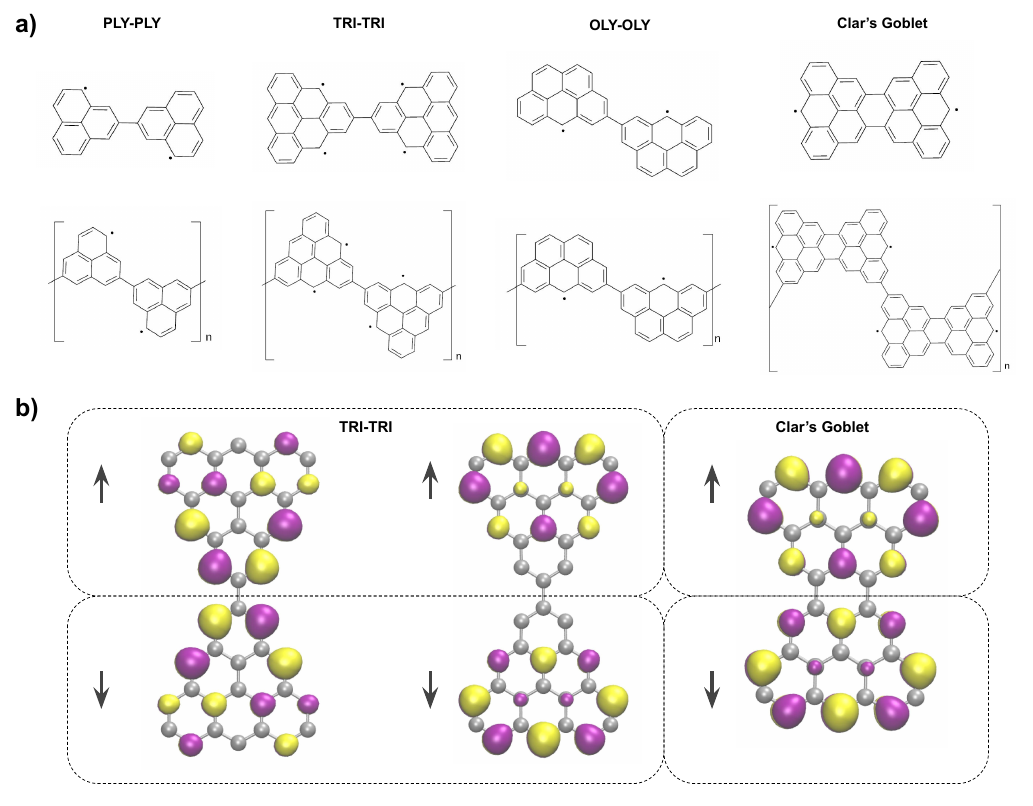}
    \caption{Representative open-shell nanographene structures and spin-resolved molecular orbitals. (a) Lewis structures of PLY–PLY, TRI–TRI, OLY–OLY, and Clar's goblet (upper row), together with schematic representations of their corresponding extended structures (lower row). Dots indicate nominal radical centers. (b) Selected spin-resolved Kohn–Sham molecular orbitals obtained from DFT calculations for TRI–TRI and Clar's goblet. Upward and downward arrows indicate the spin-up ($\alpha$) and spin-down ($\beta$) channels, respectively, while yellow and purple isosurfaces represent opposite phases of the molecular orbitals.}
    \label{fig:nanographenes}
\end{figure}

\begin{figure}[ht]
    \centering
    \includegraphics[width=1\linewidth]{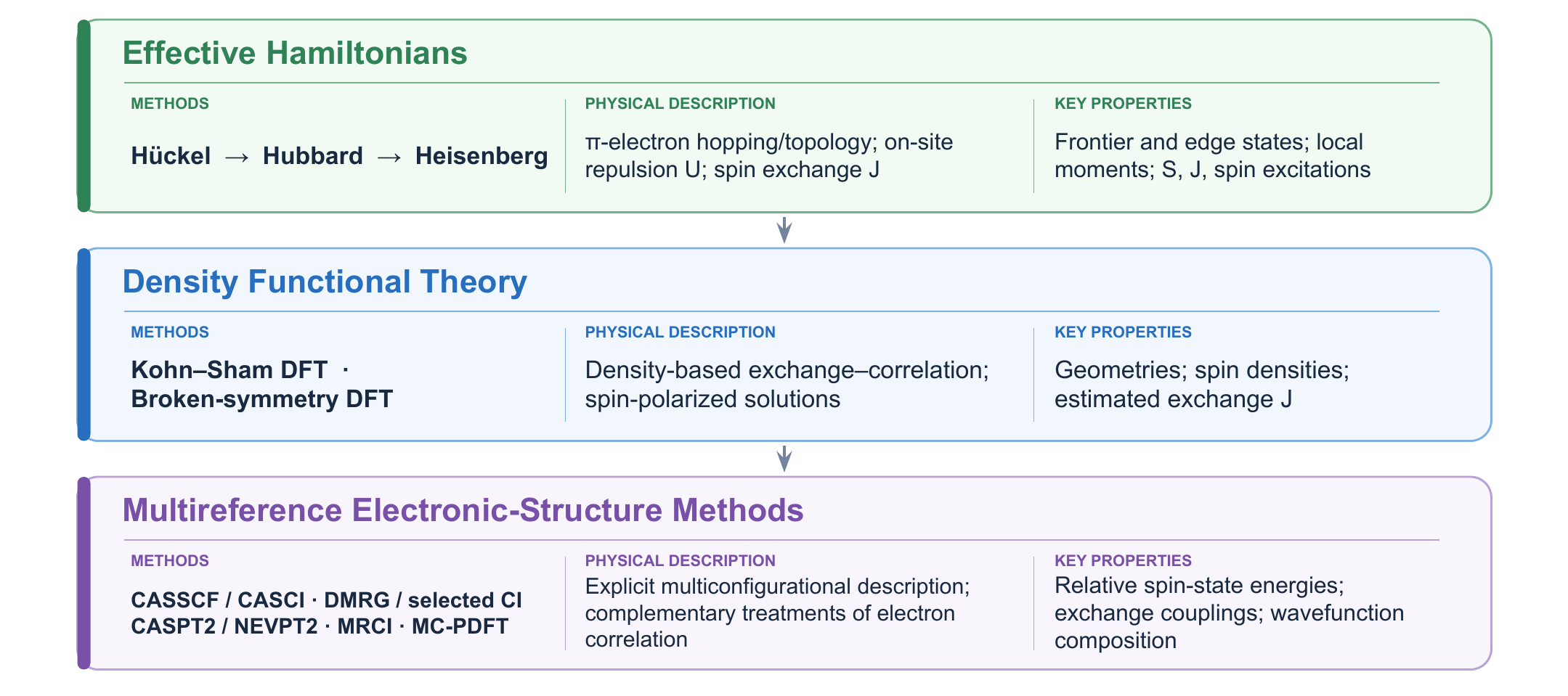}
    \caption{Schematic overview of the main theoretical approaches employed to investigate the electronic structure and magnetic properties of open-shell nanographenes. The diagram illustrates the progression from effective Hamiltonians (Hückel, Hubbard, and Heisenberg models) to density functional theory (DFT) and multireference electronic-structure methods. For each class of approaches, the underlying physical description and the key properties accessible through the calculations are highlighted.}
    \label{fig:methods}
\end{figure}

The computational investigation of the electronic structure of magnetic nanographenes requires theoretical approaches capable of capturing the delicate interplay between electron delocalization and strong electron-electron correlation \cite{Yu2023,Yu2023b,Yu2024,Yu2024c,Yu2026}. Unlike pristine, infinite graphene, finite nanographenes exhibit quantum confinement and edge-dependent phenomena. The finite size, quantum confinement, and the emergence of localized frontier states enhance the importance of strong electron correlation relative to extended graphene, giving rise to nearly degenerate frontier orbitals and multiconfigurational electronic states that cannot, in principle, be accurately described within independent-particle approximations. The resulting competition between electron delocalization and electron correlation governs many of the key properties that make nanographenes attractive for quantum technologies, including their magnetic ground state, spin density distribution, spin gap, exchange interactions, and response to external perturbations. Representative examples of open-shell nanographenes, together with selected spin-resolved molecular orbitals illustrating their magnetic electronic structure, are shown in Fig.~\ref{fig:nanographenes}.
The accurate prediction of these quantities therefore represents the central challenge of the theoretical description of open-shell nanographenes \cite{yazyev2010emergence,song2021surface,das2016polyradical} and provides the common thread linking the different computational approaches discussed throughout this section. An overview of the main theoretical approaches discussed in the following sections is presented in Fig.~\ref{fig:methods}, highlighting their underlying physical descriptions and the key properties they provide access to.

\subsection{Effective Hamiltonians for Magnetic Nanographenes}
Rather than starting from the full many-electron Schrödinger equation, the essential low-energy physics of $\pi$-conjugated nanographenes can first be understood in terms of effective lattice Hamiltonians defined on the carbon $p_z$ orbitals. These models are naturally expressed in second quantization, where the electronic structure is described through the occupation of localized orbitals on the honeycomb lattice. This representation makes it possible to build a transparent hierarchy of models: the Hückel model describes electron delocalization within the $\pi$ network, while the Hubbard model introduces local electron-electron repulsion and therefore the minimal ingredients required to describe correlation-driven magnetism. In this framework, topological concepts such as sublattice imbalance, non-bonding molecular orbitals, and edge states emerge naturally and provide the basis for understanding the magnetic behaviour of nanographenes.\\
The simplest member of this hierarchy is the Hückel model, corresponding to the nearest-neighbour $\pi$-orbital tight-binding (TB) approximation commonly employed for graphene and related nanostructures \cite{zum1931quantentheoretische,reich2002tight,liu2025electronic,kuc2010structural}. This model provides a minimal description of the $\pi$-electron system in conjugated hydrocarbons. It assumes one $p_z$ orbital per carbon atom and restricts the electronic motion to nearest-neighbour hopping, while neglecting explicit electron-electron interactions. Within the second-quantization formalism, the corresponding Hamiltonian can be written as
\begin{equation}
\hat{H}_{\mathrm{H}} =
-t \sum_{\langle i,j\rangle,\sigma}
\left(
\hat{c}^{\dagger}_{i\sigma}\hat{c}_{j\sigma}
+
\mathrm{h.c.}
\right),
\end{equation}
where the hopping integral $t$ quantifies the electronic coupling between neighbouring $p_z$ orbitals and therefore controls the degree of electron delocalization across the $\pi$ network. In pristine graphene, $t$ is typically on the order of $2.7$~eV \cite{reich2002tight}. Although remarkably simple, the Hückel model captures several qualitative features of conjugated carbon systems, including the ordering of frontier molecular orbitals, the electronic topology of the honeycomb lattice, and the emergence of edge-localized states. However, because electron-electron interactions are neglected, it cannot describe the correlated magnetic ground states characteristic of open-shell nanographenes.\\
Solving the Hückel Hamiltonian already reveals several of the key electronic features that distinguish finite nanographenes from extended graphene. Since graphene and many nanographenes are alternant hydrocarbons, their $\pi$-electron network forms a bipartite honeycomb lattice that can be partitioned into two sublattices, conventionally denoted A and B, with nearest-neighbour hopping occurring only between atoms belonging to opposite sublattices. This seemingly simple topological property has profound consequences for the electronic structure of finite systems \cite{longuet1950some}.

The origin of these features was first rationalized by Longuet-Higgins through the theory of alternant hydrocarbons \cite{longuet1950some}. In particular, he showed that finite bipartite $\pi$-conjugated systems may support non-bonding molecular orbitals lying at the Hückel non-bonding energy. These orbitals satisfy the so-called zero-sum rule, whereby the coefficients on neighbouring atoms cancel exactly. In finite nanographenes containing zigzag edges, this condition naturally leads to frontier orbitals that become strongly localized at the molecular boundaries while rapidly decaying towards the interior of the $\pi$ network \cite{nakada1996edge}. Importantly, this localization arises entirely from the topology of the lattice and the connectivity of the $\pi$ orbitals, without invoking electron-electron interactions. These frontier states play a central role in the electronic structure of open-shell nanographenes, as they provide the orbitals in which electron-electron interactions become dominant.\\
While the Hückel model successfully captures the topology of the $\pi$-electron network and the emergence of edge-localized frontier orbitals, it completely neglects the Coulomb repulsion between electrons. The simplest interacting extension is provided by the Hubbard model \cite{hubbard1963electron}, which introduces an energetic penalty for the double occupation of a localized $p_z$ orbital. The corresponding Hamiltonian can be expressed as

\begin{equation}
\hat{H}_{\mathrm{Hub}}
=
\hat{H}_{\mathrm{H}}
+
U\sum_i
\hat{n}_{i\uparrow}\hat{n}_{i\downarrow},
\end{equation}

where $U$ denotes the on-site Coulomb repulsion and $\hat{n}_{i\sigma}=\hat{c}^{\dagger}_{i\sigma}\hat{c}_{i\sigma}$ is the number operator for electrons of spin $\sigma$ occupying site $i$. The Hubbard Hamiltonian therefore describes the competition between two opposing physical mechanisms: the hopping term, which favors electron delocalization over the entire $\pi$ network, and the Coulomb interaction, which penalizes double occupation and promotes electron localization. The balance between these two contributions, commonly characterized by the ratio $U/t$, determines the electronic regime of the system, ranging from weakly correlated itinerant electrons to strongly correlated localized magnetic moments.\\
The consequences of the Hubbard model become particularly evident in the presence of edge-localized frontier orbitals. As these states are spatially confined, the effective kinetic energy associated with electron hopping is significantly reduced, while the on-site Coulomb repulsion remains essentially unchanged. Consequently, interactions can favor singly occupied frontier states and the formation of localized magnetic moments. Their mutual alignment, however, depends on the molecular topology and competing exchange mechanisms, including antiferromagnetic superexchange, rather than following from Hund's rule alone. In nanographenes, this mechanism is responsible for the emergence of magnetism at zigzag edges and other topologically frustrated motifs, despite the fact that the material is composed exclusively of non-magnetic $sp^2$ carbon atoms \cite{son2006energy,fernandez2007magnetism}.\\
Unfortunately, the Hubbard Hamiltonian cannot be solved analytically except for a few special cases, and exact numerical solutions rapidly become intractable as the system size increases. Consequently, a variety of approximate approaches have been developed. Among these, the mean-field Hubbard (MFH) approximation has become one of the most widely employed tools for investigating large graphene nanostructures. By replacing the two-electron interaction with an effective one-electron potential that is determined self-consistently, the MFH approach provides an efficient description of spin-polarized ground states while retaining the essential competition between electron delocalization and local Coulomb repulsion. Despite its simplicity, it successfully reproduces many qualitative features of magnetic nanographenes, including the localization of spin density at zigzag edges and the magnetic coupling between distant radical centers.\\
Although the Hubbard Hamiltonian captures the essential mechanism responsible for correlation-driven magnetism, obtaining its exact solution remains challenging for all but the smallest systems. Nevertheless, one of its most remarkable properties was established rigorously by Lieb \cite{lieb1989two}, who proved that the total spin of the ground state of the half-filled repulsive Hubbard model on a bipartite lattice is determined exclusively by the imbalance between the two sublattices,

\begin{equation}
S=\frac{1}{2}\left|N_A-N_B\right|,
\end{equation}

where $N_A$ and $N_B$ denote the number of carbon atoms belonging to sublattices A and B, respectively. Remarkably, this result is independent of the magnitude of the on-site Coulomb interaction, provided that $U>0$. Consequently, nanographenes with balanced sublattices ($N_A=N_B$) possess a singlet ($S=0$) ground state, whereas structures exhibiting a global sublattice imbalance, such as triangulenes and other bipartite nanographenes with uncompensated sublattices, are predicted to host nonzero-spin ground states \cite{pavlivcek2017synthesis,wang2009topological}. Within these assumptions, Lieb's theorem therefore provides a rigorous topological criterion for predicting the total spin directly from lattice structure, making it one of the most powerful conceptual tools for the rational design of magnetic graphene nanostructures.\\
While the Hubbard model explicitly describes the electronic degrees of freedom responsible for magnetism, many experimentally relevant properties depend only on the interaction between the resulting localized magnetic moments. In the strongly correlated limit ($U \gg t$) when well-localized magnetic moments are formed, charge fluctuations become energetically unfavorable and the low-energy physics can be projected onto an effective spin Hamiltonian. In this regime, the Hubbard model reduces to the Heisenberg Hamiltonian \cite{anderson1959new,takahashi1977half,macdonald1988t},

\begin{equation}
\hat{H}_{\mathrm{Heis}}
=
\sum_{i<j}
J_{ij}
\hat{\mathbf{S}}_i\cdot\hat{\mathbf{S}}_j,
\label{H_heis}
\end{equation}

where $\hat{\mathbf{S}}_i$ is the spin operator associated with the localized magnetic moment at site $i$, and $J_{ij}$ is the magnetic exchange coupling between sites $i$ and $j$. Within this convention, positive values of $J_{ij}$ correspond to antiferromagnetic coupling, whereas negative values favor ferromagnetic alignment. For the half-filled two-site Hubbard model, second-order perturbation theory yields \cite{anderson1950antiferromagnetism,macdonald1988t}

\begin{equation}
J=\frac{4t^2}{U},
\end{equation}

demonstrating that the magnetic interaction originates from the competition between electron hopping and on-site Coulomb repulsion. Larger hopping amplitudes favor stronger magnetic exchange, whereas increasing electron-electron repulsion suppresses charge fluctuations and weakens the effective coupling between localized spins. Consequently, the Heisenberg Hamiltonian provides an intuitive and computationally efficient description of the low-energy magnetic properties of nanographenes, while retaining the essential physics encoded in the underlying Hubbard model.\\
The hierarchy of effective Hamiltonians presented above provides a remarkably successful framework for understanding the physical origin of magnetism in nanographenes, from the emergence of edge-localized frontier orbitals to the formation and interaction of localized magnetic moments. Nevertheless, these models remain approximate descriptions of the electronic structure. Their predictive power ultimately relies on the choice of a limited number of model parameters, such as the hopping integral ($t$), the on-site Coulomb repulsion ($U$), or the exchange coupling ($J$), which are generally not known \textit{a priori} and may vary significantly with the molecular topology, chemical functionalization, and environment \cite{mishra2020collective,mishra2021large,Biswas2023}.\\
For quantitative predictions, it is therefore necessary to move beyond effective Hamiltonians and explicitly solve, albeit approximately, the full many-electron electronic Hamiltonian. Over the past decades, a wide range of \textit{ab initio} electronic-structure methods have been developed to address this challenge, offering different balances between computational cost and accuracy. The following section reviews the principal computational approaches currently employed to investigate the electronic structure and magnetic properties of open-shell nanographenes.

\section{Ab Initio Electronic Structure Methods}\label{sec:abinitio}

Unlike the effective Hamiltonians discussed in the previous section, \textit{ab initio} electronic-structure methods seek to describe nanographenes starting directly from the fundamental interactions between electrons and atomic nuclei. Within the Born--Oppenheimer approximation, the nuclei are treated as fixed while the electronic wavefunction is obtained by solving the time-independent electronic Schrödinger equation,

\begin{equation}
\hat{H}_{\mathrm{el}}\Psi(\mathbf{x}_1,\ldots,\mathbf{x}_N)
=
E
\Psi(\mathbf{x}_1,\ldots,\mathbf{x}_N),
\end{equation}

where $\Psi$ is the many-electron wavefunction and $\hat{H}_{\mathrm{el}}$ is the electronic Hamiltonian,

\begin{equation}
\hat{H}_{\mathrm{el}}
=
-\sum_i\frac{1}{2}\nabla_i^2
-\sum_{iA}\frac{Z_A}{r_{iA}}
+\sum_{i<j}\frac{1}{r_{ij}}.
\end{equation}

Here, atomic units are employed, the constant nuclear--nuclear repulsion energy is omitted, and the three terms represent the electronic kinetic energy, the electron--nucleus Coulomb attraction, and the electron--electron Coulomb repulsion, respectively. In contrast to the effective Hamiltonians introduced in the previous section, no empirical parameters are introduced and all interactions are treated explicitly. The exact solution of this equation would provide a complete description of the electronic structure and magnetic properties of a nanographene.\\
Unfortunately, the electron--electron interaction couples the motion of all electrons, making the many-electron Schrödinger equation impossible to solve exactly except for the smallest systems. Consequently, the development of electronic-structure theory has largely consisted of devising increasingly sophisticated approximations that balance computational efficiency and accuracy. These methods differ primarily in the way they approximate electron correlation, which plays a particularly important role in open-shell nanographenes owing to the presence of nearly degenerate frontier orbitals and multiconfigurational electronic states. Accordingly, the following sections review the principal computational approaches employed for these systems, ranging from conventional density functional theory to multireference wavefunction methods, many-body perturbation theory, and emerging approaches combining multireference wavefunctions with density functional approximations.

\subsection{Density Functional Theory}
Although effective Hamiltonians provide a powerful conceptual framework for understanding the origin of magnetism in nanographenes, quantitative predictions generally require an explicit solution of the many-electron electronic Hamiltonian. Among the available \textit{ab initio} approaches, Density Functional Theory (DFT) has emerged as the workhorse method for investigating the structural, electronic, and magnetic properties of graphene-based materials owing to its favorable balance between computational cost and predictive accuracy.\\
Historically, the first practical electronic-structure methods were based on Hartree--Fock (HF) theory, in which the many-electron wavefunction is approximated by a single Slater determinant. Within this framework, each electron moves independently in the average field generated by all other electrons, providing an exact description of exchange but neglecting electron correlation. While HF often constitutes an excellent starting point for weakly correlated systems, its inability to account for correlation effects severely limits its applicability to open-shell nanographenes, where several nearly degenerate electronic configurations contribute to the ground state.\\
DFT circumvents this limitation by reformulating the electronic-structure problem in terms of the ground-state electron density rather than the many-electron wavefunction. The Hohenberg--Kohn theorems establish that the ground-state density uniquely determines all observables of the system, while the Kohn--Sham formalism maps the interacting many-electron problem onto an auxiliary system of non-interacting electrons moving in an effective potential. Within this approach, all many-body effects are formally contained in the exchange--correlation functional, whose approximation ultimately determines the accuracy of a DFT calculation \cite{hohenberg1964inhomogeneous,kohn1965self}.\\
The computational efficiency of Kohn--Sham DFT has made it the method of choice for studying magnetic nanographenes ranging from finite molecular systems to graphene nanoribbons and extended periodic structures. In particular, DFT provides reliable equilibrium geometries, spin-density distributions, magnetic anisotropies, and approximate magnetic exchange couplings at a computational cost that permits calculations on systems containing several hundreds of atoms \cite{son2006energy,ruffieux2016surface}. These characteristics have also established DFT as the primary engine for high-throughput computational screening and data generation for machine-learning applications.\\

Despite these advantages, conventional Kohn--Sham DFT calculations rely on an auxiliary single-determinant representation, and commonly employed exchange--correlation functionals may encounter significant difficulties in describing systems with pronounced multiconfigurational character, as frequently occurs in open-shell nanographenes possessing nearly degenerate frontier orbitals. In these situations, different computational strategies have been developed depending on the physical problem under investigation. The first consists in deliberately breaking the spin symmetry of the Kohn--Sham determinant, giving rise to the Broken-Symmetry DFT approach discussed below. Alternatively, genuinely multiconfigurational wavefunction methods, such as CASSCF and its extensions, explicitly account for the simultaneous contribution of multiple electronic configurations and will be discussed in the following subsection.

\subsection{Broken-Symmetry Density Functional Theory}

One of the principal limitations of conventional Kohn--Sham DFT arises when describing antiferromagnetically coupled open-shell systems. In these molecules, the exact electronic ground state is often an open-shell singlet, whose wavefunction cannot be represented by a single Slater determinant. Consequently, a spin-restricted single-determinant description may fail to represent localized antiferromagnetically coupled moments, whereas an unrestricted calculation may stabilize either a high-spin or a broken-symmetry solution, depending on its initial guess and the system.\\
To overcome this limitation, Noodleman introduced the Broken-Symmetry (BS) approach \cite{noodleman1981valence,noodleman1986ligand}, in which the spin symmetry of the Kohn--Sham determinant is deliberately relaxed. Rather than enforcing a pure spin eigenfunction, the electronic optimization is allowed to converge towards a state in which the localized magnetic moments adopt opposite spin orientations. Although the resulting wavefunction is no longer an eigenfunction of the total spin operator $\hat{S}^2$, it often provides a realistic description of the spin density distribution and constitutes an effective approximation to the antiferromagnetically coupled electronic state.\\
The broken-symmetry solution should not be interpreted as the exact singlet wavefunction, but rather as a spin-contaminated state that mixes different spin multiplicities. Nevertheless, because it correctly reproduces the localization of the magnetic moments, it has become one of the most widely employed computational approaches for estimating magnetic exchange interactions in molecular magnets and graphene-based open-shell systems. Since the BS solution is not a spin eigenfunction, the exchange coupling $J$ cannot be obtained directly from the energy difference between the high-spin and broken-symmetry solutions. Instead, several projection schemes have been proposed to map the BS energies onto the Heisenberg spin Hamiltonian in Eq. \ref{H_heis}.\\
The existence of multiple projection schemes reflects the fact that the broken-symmetry determinant is not uniquely related to a pure spin state. Different approaches therefore adopt different assumptions regarding the projection of the spin-contaminated solution onto the eigenstates of the Heisenberg Hamiltonian. Although all these methods seek to estimate the same physical quantity, namely the magnetic exchange coupling $J$, they differ in their theoretical justification, numerical implementation, and range of applicability \cite{bencini2009few,ruiz1999broken,ruiz2005calculation}.\\
The three projection schemes most commonly employed in the literature are those proposed by Noodleman \cite{noodleman1981valence}, Ruiz \cite{ruiz2005calculation}, and Yamaguchi \cite{yamaguchi1988ab,shoji2006general}. For clarity, the following discussion focuses on the interaction between two equivalent localized spin centers $S_1 = S_2$. This corresponds to the simplest realization of the more general Heisenberg Hamiltonian introduced in Eq. \ref{H_heis}, i.e. $\hat{H}_{\mathrm{Heis}}=J\hat{\mathbf S}_1\cdot\hat{\mathbf S}_2$.\\
In the weak-coupling limit, the original spin-projection scheme proposed by Noodleman assumes a small overlap between the localized magnetic orbitals, leading to

\begin{equation}
J_{\mathrm{N}}
=
\frac{2(E_{\mathrm{HS}}-E_{\mathrm{BS}})}
{S_{\max}^{2}},
\end{equation}
where $S_{\max}$ is the total spin quantum number of the high-spin state.\\
In contrast, Ruiz and co-workers proposed a strong-coupling approximation via a direct mapping between the electronic energies and the eigenvalues of the Heisenberg spin Hamiltonian. Within this approximation, the exchange coupling is given by

\begin{equation}
J_{\mathrm{R}}
=
\frac{2(E_{\mathrm{HS}}-E_{\mathrm{BS}})}
{S_{\max}(S_{\max}+1)}.
\end{equation}
A more general approximate spin-projection expression, widely employed in computational chemistry, was proposed by Yamaguchi and co-workers:
\begin{equation}
J_{\mathrm{Y}}
=
\frac{2(E_{\mathrm{HS}}-E_{\mathrm{BS}})}
{\langle\hat S^2\rangle_{\mathrm{HS}}
-
\langle\hat S^2\rangle_{\mathrm{BS}}},
\end{equation}
which explicitly accounts for the spin contamination of the broken-symmetry solution through the expectation value of the total spin operator.\\
The Noodleman ($J_{\mathrm{N}}$) and Ruiz ($J_{\mathrm{R}}$) expressions therefore correspond to different assumptions about the relationship between the BS determinant and the low-spin state: the former is associated with weakly overlapping magnetic orbitals, while the latter assumes that the BS energy closely approximates the energy of a pure low-spin state. By using the calculated $\langle\hat S^2\rangle$ values, the Yamaguchi expression offers a more flexible approximation across these regimes \cite{ruiz1999broken,bencini2009few,singh2021magnetic}. These mappings need not give identical values even for localized spin centres: for two $S=1/2$ moments, the Noodleman and Ruiz denominators differ by a factor of two. Their validity also depends on the adequacy of the Heisenberg mapping and the quality of the underlying electronic-structure description. In particular, spin contamination, orbital delocalization, and multiconfigurational character can limit the reliability of the extracted $J$ values; comparisons with multireference calculations or experiment therefore remain desirable \cite{iwahara2025breakdown}.

\subsection{Multireference Wavefunction Methods}

Although Broken-Symmetry DFT provides an efficient framework for estimating magnetic exchange interactions in large nanographenes, it remains fundamentally based on a single-determinant description of the electronic structure. Consequently, the broken-symmetry solution only approximates the true multiconfigurational wavefunction through spin symmetry breaking. Whenever several electronic configurations contribute significantly to the ground state, a more rigorous treatment requires methods that explicitly optimize a multideterminantal wavefunction \cite{Yu2023,Yu2023b}.\\
The Complete Active Space Self-Consistent Field (CASSCF) method addresses this problem \cite{roos1980complete} by explicitly constructing the wavefunction as a linear combination of all possible electronic configurations generated within a selected set of chemically relevant molecular orbitals, known as the active space. Electrons occupying inactive orbitals remain doubly occupied throughout the calculation, whereas virtual orbitals outside the active space remain empty. Only the active electrons are allowed to distribute freely among the active orbitals, thereby providing a balanced description of the multiple electronic configurations responsible for static correlation.\\
The notation CAS($n,m$) indicates that $n$ active electrons are distributed among $m$ active orbitals by considering all possible electronic configurations compatible with the chosen spin symmetry. The quality of a CASSCF calculation therefore depends critically on the selection of the active space, which should include all orbitals participating in bond breaking, magnetic exchange, or near-degenerate electronic configurations. For magnetic nanographenes, the active space is typically constructed from the frontier $\pi$ orbitals responsible for the open-shell character and magnetic coupling \cite{das2016polyradical}.\\
In principle, extending the active space towards the complete $\pi$-electron manifold offers a systematic way to assess the convergence of a multiconfigurational description of a nanographene. Such a treatment would simultaneously account for the static correlation associated with nearly degenerate frontier orbitals and the delocalized nature of the $\pi$ electrons over the entire conjugated framework. Unfortunately, the dimension of the complete active-space expansion increases combinatorially with the number of active electrons and orbitals, making conventional CASSCF calculations rapidly intractable as the molecular size grows \cite{hachmann2007radical,levine2020casscf}. Consequently, practical calculations require a compromise between computational feasibility and physical completeness, whereby only a subset of the most chemically relevant orbitals is included in the active space.\\
This limitation largely explains the complementary roles played by Broken-Symmetry DFT and multireference wavefunction methods in the theoretical investigation of magnetic nanographenes. Whereas BS-DFT enables routine calculations on systems containing hundreds of atoms, conventional CASSCF calculations are generally restricted to comparatively small active spaces. Consequently, multireference methods are frequently employed as high-accuracy benchmark calculations for representative systems, while BS-DFT remains the practical choice for large-scale screening and realistic nanographene architectures.\\
Although CASSCF provides an accurate treatment of static correlation, it neglects most of the dynamic correlation arising from the instantaneous motion of electrons outside the active space. This contribution is essential for obtaining quantitatively reliable magnetic exchange couplings, excitation energies, and relative energetics. Consequently, CASSCF is commonly combined with multireference perturbation theories, among which Complete Active Space Second-Order Perturbation Theory (CASPT2) and $N$-Electron Valence State Second-Order Perturbation Theory (NEVPT2) have become the most widely employed \cite{andersson1990second,andersson1992second,angeli2001introduction,angeli2002n}. While both methods provide an efficient description of dynamic correlation, NEVPT2 has become particularly popular owing to its intruder-state-free formulation, size consistency, and parameter-free character \cite{angeli2002n}.\\
Multireference configuration interaction (MRCI) provides a variational alternative for recovering correlation beyond the active space. Starting from a multiconfigurational reference, MRCI expands the wavefunction with excited configurations and can describe states with substantial multireference character, including their relative energies. Internally contracted formulations mitigate the computational cost, although truncated MRCI is not generally size-extensive and becomes expensive for large nanographenes \cite{werner1988efficient}.\\
Despite these advances, the exponential scaling associated with the complete active-space expansion remains the principal bottleneck of multireference electronic-structure methods. This challenge is especially severe for nanographenes, where a quantitatively converged description may require active spaces comprising several tens of correlated $\pi$ orbitals. To overcome this limitation, two complementary classes of methods have become particularly important.\\
The first is represented by the Density Matrix Renormalization Group (DMRG), which replaces the full configuration-interaction expansion within the active space by a matrix-product-state representation of the wavefunction \cite{white1992density,chan2011density}. Combined with orbital optimization (DMRG-SCF or DMRG-CASSCF), this approach can routinely treat active spaces far beyond the reach of conventional CASSCF while preserving a variational description of static correlation \cite{hachmann2007radical,ghosh2008orbital,sharma2012spin,chan2011density}. \\
A conceptually different strategy is followed by selected configuration interaction (sCI) methods, which approximate the complete active-space wavefunction by iteratively selecting only the most important electronic configurations. Among the most recent developments, the Iterative Configuration Expansion (ICE) method combines perturbative determinant selection with spin-adapted configuration state functions, allowing compact multiconfigurational wavefunctions to be constructed while rigorously preserving spin symmetry \cite{chilkuri2021comparison,chilkuri2021comparisonII,guo2025approximations}. \\
The development of DMRG and ICE represents a major advance for the theoretical investigation of magnetic nanographenes. Rather than replacing conventional CASSCF, these approaches extend its applicability to significantly larger active spaces, enabling systematic studies of the convergence of magnetic properties with respect to the active-space size. 
Another promising strategy is to combine multiconfigurational wavefunctions with density-functional approximations. In multiconfiguration pair-density functional theory (MC-PDFT), the electron density and on-top pair density obtained from a multiconfigurational reference are used with a density functional to recover additional correlation at a cost generally lower than that of conventional post-CASSCF treatments \cite{li2014multiconfiguration}. Recent applications of range-separated MC-PDFT to triangulene, heptauthrene, diphenalenyl, and Clar's goblet illustrate its potential for describing the magnetic coupling of open-shell nanographenes, while also highlighting the importance of the active space and the selected functional \cite{Rodrigues2025}.

\subsection{Quasiparticle Renormalization and Substrate Screening: Beyond Ground-State DFT}

A fundamental limitation of standard Kohn-Sham DFT arises when researchers attempt to directly compare computed orbital energies to experimental Scanning Tunneling Spectroscopy (STS) measurements. Fundamentally, Kohn-Sham eigenvalues are mathematically auxiliary constructs; they do not represent true quasiparticle excitation energies. While STS physically injects or extracts an electron—measuring the true fundamental gap defined by the Ionization Potential ($IP$) and Electron Affinity ($EA$)—ground-state DFT inherently struggles to capture the localized energetic penalty of these charged excitations. 

This discrepancy becomes more important when the nanographene is adsorbed onto a metallic substrate, as in most part of experiments. When an electron is injected into (or removed from) the molecule via the STM tip, the highly polarizable two-dimensional electron gas of the underlying metal reacts instantaneously. The substrate electrons redistribute to create a dynamic image charge that electrostatically screens the added carrier. This macroscopic dielectric response stabilizes the transient cationic or anionic states of the molecule. Consequently, the quasiparticle Highest Occupied Molecular Orbital (HOMO) is pushed up in energy, and the Lowest Unoccupied Molecular Orbital (LUMO) is pulled down. This screening effect can drastically shrink the fundamental transport gap—often by several electron volts compared to the gas-phase calculation—leading to a massive overestimation of the gap if standard gas-phase DFT is used as the sole point of comparison.

To computationally capture this quasiparticle renormalization and directly simulate STS conductance spectra, approaches that go beyond ground-state density functionals must be employed, \textit{i.e.} Many-Body Perturbation Theory (MBPT), most notably the $GW$ approximation. The primary objective of the $GW$ approximation is to correct the systematic ``band gap problem'' inherent in standard density functional theory. While Kohn-Sham DFT is a ground-state theory that treats electrons as moving in an averaged local potential, it fails to account for the discrete nature of electron addition and removal. Specifically, $GW$ provides the necessary quasiparticle corrections to account for the energy of the ``hole'' left behind during photoemission or the ``added electron'' in inverse photoemission. Here, $G$ represents the one-particle Green's function (describing the propagation of the injected electron or hole), and $W$ is the dynamically screened Coulomb interaction. By replacing the static exchange-correlation functional with a dynamic self-energy, $GW$ eliminates the unphysical self-interaction of electrons and correctly captures the many-body polarization of the environment, thereby shifting the HOMO and LUMO levels to match true experimental ionization potentials and electron affinities. In graphene nanoribbons, $GW$ corrections to conventional DFT gaps can be substantial, and dielectric screening by a supporting substrate can further renormalize these energies \cite{hedin1965new,yang2007quasiparticle,Kharche2016}. 

In the single-shot $G_0W_0$ approach, the self-energy is calculated using the initial Kohn-Sham eigenvalues and wavefunctions. $G_0W_0$ is widely utilized for its balance of accuracy and cost, even if it suffers from a significant ``starting point dependence'', since the final quasiparticle energies remain sensitive to the choice of the underlying exchange-correlation functional.  By explicitly calculating the frequency-dependent dielectric matrix of the combined molecule-metal system, the $G_0W_0$ approach natively captures the substrate-induced image-charge screening, yielding theoretical STS gaps that achieve quantitative agreement with experiment \cite{Wilhelm2018,Biswas2023}.

However, the computational cost of solving the full frequency-dependent $W$ scales immensely with system size ($\mathcal{O}(N^4)$ to $\mathcal{O}(N^6)$), making explicit $G_0W_0$ calculations on massive nanographene-metal slabs practically prohibitive. To circumvent this, two highly effective approximations can be reliably employed:

\vspace{0.5em}
\noindent \textbf{1. The Static COHSEX Approximation.}
To reduce the computational burden, the dynamic (frequency-dependent) self-energy can be approximated in the static limit ($\omega \rightarrow 0$) using the Coulomb Hole and Screened Exchange (COHSEX) approximation \cite{Kharche2016}. This method physically separates the self-energy into two distinct terms: the Screened Exchange (SEX), which accounts for the Pauli exclusion principle modulated by the dielectric medium, and the Coulomb Hole (COH), which models the classical electrostatic stabilization of the added charge. By removing the dynamic frequency dependence, COHSEX vastly reduces computational overhead while successfully recovering the bulk of the substrate screening effects.

\vspace{0.5em}
\noindent \textbf{2. Image-Charge Corrected DFT (DFT+IC).}
For larger systems where even COHSEX is intractable, semi-classical image-charge corrections \cite{Urgel2019,Xu2019} can be employed. In this approach, a standard, computationally cheap gas-phase DFT calculation is performed, and the resulting Kohn-Sham eigenvalues are perturbatively corrected using a classical electrostatic model. By defining the vertical distance ($z$) between the molecular plane and the metal's image plane, and assigning a bulk dielectric constant to the substrate, a physically motivated $\Delta E \propto -1/(4z)$ energy correction is applied directly to the HOMO and LUMO levels. This highly scalable DFT+IC approach provides a remarkably accurate, computationally trivial map of the renormalized STS gap for extended 2D architectures.

\vspace{0.5em}
\noindent \textbf{3. Dyson Orbitals.} As an alternative approach, the interpretation of STS for magnetic nanographenes has recently shifted from traditional one-electron molecular orbital (MO) theory toward a more robust multireference framework based on Dyson orbitals \cite{Kumar2025}. 

Dyson orbitals formally represent the overlap (transition amplitude) between the many-body $N$- and $(N \pm 1)$-electron wavefunctions. Because they capture the change in electron correlation upon ionization or electron attachment, Dyson orbitals directly dictate the actual probabilities of the electronic transitions detected in an STM junction. 

Applications to diverse nanographene systems have demonstrated superior agreement with experimental $dI/dV$ maps compared to pure DFT methods, particularly in capturing the subtle spectral weights of frustrated spin states. However, a significant frontier remains: while these multireference methods provide a grounded description of the molecular many-body states, they still struggle to explicitly incorporate the semi-infinite metallic surface, often necessitating image-charge corrections or embedding schemes.

\section{The Substrate-Molecule Interface: Environmental Effects and Reactivity}

	\subsection{Chemisorption vs Physisorption: Substrate Selectivity and Magnetic States}\label{sec:chemi_physi}
	
	While isolated vacuum models provide foundational insights into the intrinsic quantum mechanics of nanographenes, they represent a physically incomplete picture. In modern experimental realization—typically via ultra-high vacuum (UHV) on-surface synthesis—the nanographene is strongly coupled to a metallic or dielectric substrate (e.g., Au(111), Ag(111), Cu(111)). The substrate fundamentally dictates the reaction pathways, structurally deforms the molecule, and radically alters the electronic and magnetic ground states. Consequently, computational investigations must explicitly incorporate the surface to capture several overlapping physical phenomena.
	
	The determination of the equilibrium adsorption height is the most immediate consequence of the interaction between a magnetic nanographene and its supporting substrate. This height is governed by a delicate balance between long-range attractive van der Waals (vdW) forces and short-range Pauli repulsion, which arises from the overlap between the molecular $\pi$-system and the electronic states of the metal surface. This Pauli repulsion is the same physical principle exploited in non-contact atomic force microscopy (nc-AFM) with carbon monoxide (CO)-functionalized tips to achieve submolecular resolution of the carbon backbone \cite{Mishra2020}. On noble metal surfaces like Au(111), nanographenes typically adsorb at a height of approximately 3.1 \AA, indicating a regime of weak physisorption \cite{Urgel2019}.
	
	Accurately capturing this adsorption geometry computationally poses significant challenges. Standard semilocal DFT functionals drastically fail to model long-range dispersion, necessitating explicit vdW corrections to predict adsorption energies and heights accurately. While empirical pairwise schemes, such as Grimme's D3~\cite{grimme2010consistent,grimme2011effect}
    or D4~\cite{dftd4-1,dftd4-2,dftd4-periodic}, are widely used for geometry optimization
    due to their computational efficiency, they often struggle to account for the dynamic
    screening of highly delocalized metallic electrons.
     To properly resolve these interface dynamics, computational studies increasingly employ the Many-Body Dispersion (MBD) formalism or non-local vdW density functionals (e.g., vdW-DF, optB88-vdW, rVV10) \cite{Cheng2022}.
	
	
	The adsorption height is a critical parameter because it exponentially controls the degree of electronic hybridization between the molecule and the substrate. A smaller height leads to stronger interaction, which can significantly alter the molecule's intrinsic properties \cite{Mishra2020}. On metallic surfaces, image charge screening plays a dominant role, leading to a substantial renormalization of the quasiparticle gap. For instance, GW calculations have shown that the energy gap of triangulene or $\pi$-extended triangulene on Au(111) is reduced by approximately 1 eV compared to the gas phase \cite{su2019atomically}.
	
	Furthermore, proximity to the metal surface renormalizes excitation energies through Kondo exchange and reduces the effective value of the on-site Coulomb repulsion (Hubbard $U$) due to screening by conduction electrons. A clear example is seen in the phenalenyl dimer, where $U$ is estimated to be $\approx 2.5$~eV on Au(111) compared to $\approx 5.4$~eV on an insulating NaCl layer \cite{Krane2023}. Indeed, the choice of substrate severely dictates the nature of the chemical bonding and the resulting magnetic state:
	
	\begin{itemize}
		\item \textbf{Au(111)}. Usually considered relatively inert, it allows nanographenes to physisorb and preserve their intrinsic magnetic ground states, such as the triplet state of triangulene or the exchange-coupled states of $\pi$-extended triangulene. However, site-specific effects can occur; at low coverage, molecules like Clar's goblet can become pinned to the ``elbows'' of the Au(111) herringbone reconstruction. These undercoordinated sites act as reactive centers, leading to site-specific chemisorption that quenches local magnetic moments and transforms the system (e.g., from an $S=0$ antiferromagnet into an effective $S=1/2$ system) \cite{Mishra2020}.
		\item \textbf{Cu(111)}. This highly reactive surface promotes strong $d$-band hybridization and covalent bonding (chemisorption). This interaction often quenches the edge magnetism of zigzag nanographenes, turning open-shell radicals into closed-shell complexes hybridized with the substrate.
		\item \textbf{Ag(111)}. The lower work function of silver compared to gold can drastically alter charge-transfer regimes. For example, aza-triangulene donates an electron on Au(111) to become a triplet cation, whereas on Ag(111), it receives an electron from the substrate, assuming a closed-shell electronic configuration \cite{Wang2022}.
		\item \textbf{Insulating Decoupling Layers}. To recover intrinsic, ``gas-phase-like'' electronic and magnetic properties, researchers often use decoupling layers such as NaCl or Xe \cite{Krane2023}. These layers electronically isolate the nanographene from the metal, suppressing screening and hybridization effects and allowing direct observation of pure spin excitations and unrenormalized molecular orbitals.
	\end{itemize}
	
	\subsection{Catalysis and On-Surface Synthesis Kinetics}
	
	Ultimately, the underlying metal substrate acts as an highly active, 2D catalytic template that governs the entire bottom-up synthetic workflow. The fabrication of atomically precise nanographenes typically involves UHV deposition of halogenated molecular precursors, followed by thermal annealing to induce dehalogenation (Ullmann-type coupling) and subsequent cyclodehydrogenation (the Scholl reaction) \cite{grill2020covalent,clair_controlling_2019,grzybowski2020synthetic,cai_atomically_2010}.
	
	Gas-phase calculations of these reaction paths fail to capture the physics of these systems; the metal surface drastically lowers activation energy barriers and completely alters the underlying reaction mechanisms through the formation of transient, surface-adsorbate intermediates \cite{treier2011surface,bjork2011zipping}. 
	
	\subsubsection{Building the reaction network}
	
	Deciphering the multi-step mechanisms governing on-surface synthesis fundamentally relies on mapping the underlying potential energy surface (PES). From a structural and topological perspective, a reaction network is established by locating reactants, products, local minima (representing stable or metastable reaction intermediates), and first-order saddle points, which define the transition states (TS) linking these minima. Characterizing this complex landscape requires identifying not only the structural configurations of these stationary points but also the explicit connectivity that defines valid elemental reaction steps \cite{unsleber2020exploration}.
	
	To compute the energy and atomic forces required to navigate this landscape, researchers rely on a variety of methods and algorithms. Traditionally, DFT is coupled with chain-of-images approaches, such as the Nudged Elastic Band (NEB) or Climbing-Image NEB (CI-NEB) methods \cite{henkelman2000improved,henkelman2000climbing}, to locate TSs connecting two intermediates (double-ended methods), alongside tools like the dimer method \cite{henkelman1999dimer} (single-ended methods). Once a candidate TS structure is located, it is mathematically verified as a first-order saddle point possessing exactly one negative eigenvalue in its Hessian matrix \cite{schlegel2011geometry}. Subsequently, an intrinsic reaction coordinate (IRC) \cite{fukui1981path,maeda2015intrinsic} calculation is executed to confirm that the minimum-energy path physically connects the correct reactant and product minima.
	
	However, as nanographene precursors grow in molecular weight and architectural complexity, the dimensionality of the configurational space expands dramatically. Relying on human intuition and manual trial-and-error to propose intermediate configurations and construct initial path guesses for NEB calculations creates a severe bottleneck. Unbiased, manual exploration becomes practically impossible when handling highly flexible precursors, bulky polycyclic aromatic hydrocarbons, or competitive surface-adatom interactions, thereby hindering the rational, blueprint-to-fabrication design of novel covalent nanomaterials.
	
	\subsubsection{Computational reactivity studies on Ullmann reaction}
	
	We now turn to an overview of some of the computational approaches used in literature so far to manually explore the PES. We will focus in particular on the well-known Ullmann coupling. Originally was developed as a solution coupling reaction for aryl halides in the presence of copper as catalyst \cite{ullmann1901ueber}. Nowadays, the on-surface variant of Ullmann coupling is an established route for the tailored synthesis of 1D and 2D nanostructures \cite{fan2015surface,lackinger2017surface}. The reaction is extremely versatile due to the possibility of pre-defining the topology of covalent structures by the substitution pattern of the precursor. 1D structures, such as poly(para-phenylene) chains \cite{jacs20111dstructures} or graphene nano-ribbons \cite{acsnano2013}, are fabricated from two-fold halogenated building blocks, whereas for 2D networks multiply halogenated monomers are used \cite{grill2007nano,gutzler2009surface}.
	
\begin{figure}[ht]
    \centering
    \includegraphics[width=1\linewidth]{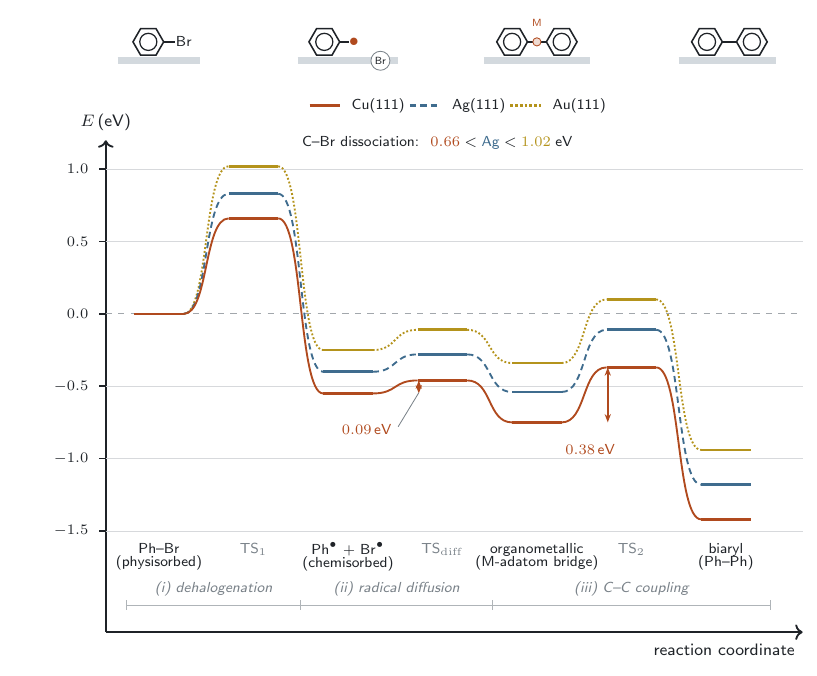}
    \caption{Substrate dependence of the on-surface Ullmann coupling: (i) dehalogenation, (ii) radical diffusion, (iii) C--C coupling through an adatom-bridged organometallic intermediate. C--Br dissociation barriers from Ref.~\cite{bjork2013mechanisms}; diffusion (0.09~eV) and coupling (0.38~eV) barriers on Cu(111) from Ref.~\cite{nguyen2011ab}. Remaining features of the Ag and Au profiles are schematic.}
    \label{fig:ullmann}
\end{figure}

	The on-surface Ullmann reaction (on coinage metal surfaces, i.e., Cu, Ag, Au) proceeds through three essential stages: (i) dehalogenation of the aryl halide precursor, (ii) diffusion of the resulting surface-stabilized radicals, and (iii) C–C coupling between two radicals to form a biaryl, as sketched in Figure~\ref{fig:ullmann}.
	
	Björk \textit{et al.} \cite{bjork2013mechanisms} carried out a DFT study of bromobenzene and iodobenzene dissociation on the (111) facets of Cu, Ag, and Au using the NEB method. They found that dehalogenation barriers follow the trend Cu(111) (0.66 eV for Br) $<$ Ag(111) $<$ Au(111) (1.02 eV for Br), establishing Au as the least reactive surface for C–X bond cleavage. The reaction is exothermic on all three metals, and the energy barriers follow a Brønsted–Evans–Polanyi relationship \cite{evans1938inertia}. For a given surface, iodobenzene dissociates more readily than bromobenzene (barrier difference ~0.3 eV), supporting hierarchic dehalogenation strategies using mixed halogen precursors.
	
	Nguyen \textit{et al.} \cite{nguyen2011ab} used DFT calculations to elucidate the atomistic mechanism of phenyl radical homocoupling on Cu(111).
	The copper substrate was simulated as a periodic slab comprised of three $8 \times 6$ Cu(111) layers separated by a 25 Å vacuum gap to isolate periodic replicas along the surface normal. To map out the reaction pathways, locate saddle points, and evaluate activation barriers for both the surface diffusion and radical coupling processes, the CI-NEB algorithm was employed with up to 16 intermediate images along the reaction coordinates. The calculations revealed that adsorbed phenyl groups exhibit strong attractive interactions and high surface mobility, migrating via an indirect diffusion path with a remarkably low activation energy of 0.09 eV. Prior to covalent bond formation, the radicals converge into a stable intermediate state mediated by localized substrate restructuring, in which a single copper atom ``pops out'' of the surface plane to bridge the reactants. Final C–C coupling to form a planar biphenyl molecule proceeds with a calculated activation barrier of 0.38 eV. Mechanistically, this reaction is electronically governed by a frontier molecular orbital alignment scheme reminiscent of Hoffmann's concepts \cite{hoffmann1988chemical}, where back-donation to the metallic substrate lifts the C–C antibonding ($\sigma^*$) state above the Fermi level.
	
	Beyond single-coupling events, Di Giovannantonio \textit{et al.} \cite{di2016mechanistic} combined fast X-ray photoelectron spectroscopy (XPS) with DFT-based NEB calculations to develop a kinetic model for the Ullmann polymerization of 1,4-dibromobenzene on Cu(110). Calculations revealed a transient state preceding C–C bond formation, from which monomers can either couple or back-diffuse to the stable organometallic phase. The kinetic model—treating nucleation (dimerization) and growth as distinct processes with different activation barriers—showed that the polymerization is diffusion-controlled and follows a nucleation-and-growth mechanism, with the nucleation barrier (1.31 ± 0.01 eV, experimentally) exceeding that for chain growth (1.09 ± 0.01 eV). The DFT-derived barriers qualitatively reproduced this trend, confirming that the stable organometallic intermediate makes dimerization more difficult than monomer addition to an existing chain.
	
	\subsubsection{Automated reaction path discovery}
	To bypass human bias and eliminate the tedious manual setup of TS guesses, the atomistic modeling of on-surface synthesis must be integrated with automated exploration frameworks. Automated path-finding algorithms methods designed for exploring reaction space can be categorized into three main groups: MD-based algorithms, heuristics-based algorithms, and algorithms focused on the study or manipulation of the PES \cite{zimmerman2018,martineznunez2026}.
	
	\paragraph{MD-based algorithms}
	
	An important portion of the automated methods developed so far are based on MD simulations. These methods sample the PES by propagating trajectories on a potential energy surface computed with a not very demanding approach: reactive force field, semiempirical, or a cheap ab initio/DFT method. Rather than relying on unguided sampling, these packages utilize tools like the ab initio nanoreactor (using virtual pistons to accelerate reactive events via GPU-driven electronic packages) or continuously monitor changes in atom-to-atom connectivity matrices. When an elementary reaction event is detected, graph theory algorithms isolate the local configuration, which is subsequently smoothed via nebterpolation and validated using IRC calculations. Examples are Ab Initio Nanoreactor \cite{wang2014discovering}, AutoMeKin \cite{amk2021}, ChemDyME \cite{shannon2021chemdyme}, ChemTraYzer \cite{chemtrayzer2015}, and ReNeGate \cite{hashemi2022renegate}. In principle these codes are developed for non-periodic systems, but the concept can be extended to periodic ones.
	Because MD simulations are computationally intensive, their application is limited to small to medium-sized systems. On the other side metadynamics (m-Dyn) \cite{laio2002escaping,barducci2011metadynamics} does sample the PES by adding biasing potentials to escape local energy wells. However, unguided m-Dyn is prone to oversimplifying complex kinetics, occasionally collapsing multi-step transformations into an artificial ``single step'' \cite{conter2026comparison}.  
	
	\paragraph{Heuristic-based algorithms}
	Heuristic methods leverage predefined chemical rules and graph-theory formalisms (treating atoms as nodes and bonds as edges) to systematically predict reaction pathways. By applying these intuitive transformation rules to an initial reactant graph, the algorithm maps a logical sequence of intermediates to the final products. A key advantage of this approach is the ability to impose strict structural constraints, preventing combinatorial explosion and making heuristic-based methods significantly more scalable for large molecular systems than unguided MD simulations. Among the codes present in literature we recall AARON \cite{guan2018aaron}, autodE \cite{duarteautode}, KinBot \cite{van2020kinbot}, RMG \cite{liu2021reaction}, ZStruct \cite{zimmerman2013automatedztruc}, YARP \cite{zhao2021simultaneouslyYARP}, and Pynta \cite{johnson2023pynta}.

	\paragraph{Topology and PES manipulation}
	Another major class of exploration frameworks focuses on the direct topological manipulation of the PES, often by applying artificial external forces to drive molecular transformations. Unlike trajectory-accelerated methods like the ab initio nanoreactor, which use forces dynamically within an MD simulation, programs such as CHEMOTON \cite{chemoton2022} and GRRM \cite{grrm2018} use force-induced and deformation-following algorithms as systematic, static tools to map local minima and saddle points across the energy landscape. Other approaches include Discrete Path Sampling (DPS) \cite{wales2002discrete,wales2004some} and Reactive Global Optimization (RGO) \cite{negreiros2012first}. DPS acts as a highly systematic framework to connect distant minima on a multidimensional PES. Instead of calculating a single minimum energy pathway, DPS automatically builds a comprehensive network of discrete pathways by iteratively applying single-ended searches to bridge gaps between known wells. This method is uniquely capable of resolving highly intricate, multi-step cascade reactions, frequently encountered during the extensive cyclodehydrogenation of bulky polycyclic aromatic hydrocarbons, where traditional NEB calculations fail to converge due to the high dimensionality of the system. On the other hand, RGO represents a powerful single-ended approach designed to systematically map the reaction network by driving the system over potential energy barriers without pre-defining the reaction coordinates. It excels at uncovering unexpected, non-intuitive structural transformations—such as rapid surface-adatom insertions or complex hydrogen shifts on crowded facets—by directly searching for maximal downward distortions or applying directional forces to escape localized minima.  
		
	\subsubsection{The machine learning interatomic potentials paradigm: bridging the accuracy-cost gap}
	
	Even with automated PES exploration algorithms, evaluating thousands of intermediate structures and Hessian matrices within a complex reaction network using conventional, first-principles DFT remains an immense computational bottleneck. To resolve this trade-off between chemical accuracy and exhaustive structural exploration, Machine Learning Interatomic Potentials (MLIPs) have emerged as a highly disruptive surrogate technology. By training neural network architectures or equivariant message-passing networks on vast datasets of DFT-computed materials properties (leveraging open large-scale databases like OC20, MPTraj, or Alexandria \cite{oc202021,chgnet2023,alexandria2026}), MLIPs learn the complex quantum-mechanical interactions between atoms. They map structures, total energies, and forces with near-DFT precision while delivering computational speedups on the order of $10^4$.
	
	In the context of atomistic modeling of nanographenes, this acceleration changes how we study synthesis kinetics:
	\begin{itemize}
		
		\item Large-Scale Reactive Molecular Dynamics: Molecular Dynamics (MD) simulations can now be run at realistic synthesis operative conditions over nanosecond timescales for large, multi-molecular surface coverages. This captures explicit entropic variations, surface-adatom diffusion kinetics, and multi-precursor crowding phenomena that static, $0\text{ K}$ DFT calculations inherently miss.
		
		\item Rapid Multi-Facet and Defect Screening: The activation barriers and competing pathways of large precursors can be screened on-the-fly across numerous step-edges, kinks, and alloy surfaces to determine exactly how surface roughness or step defects alter reaction barriers and selectivity..
		
		\item Active Learning Loops: Global optimization algorithms can actively query MLIPs to identify active catalytic sites or highly strained intermediates, automatically triggering high-level DFT calculations only when the model encounters low-confidence regions of the structural configuration space.
		
	\end{itemize}
	While ``out-of-the-box'' foundational models often require targeted fine-tuning to perfectly capture the subtle balance of localized covalent bond-breaking and delocalized vdW dispersion forces governing large aromatic molecules on metals, dedicated benchmarking platforms (such as CatBench\cite{catbench2025} or MLIPX\cite{mlipx2025}) provide unified frameworks to evaluate their readiness for surface reactivity.
	
\begin{figure}[htbp]
    \centering
    \includegraphics[width=1\linewidth]{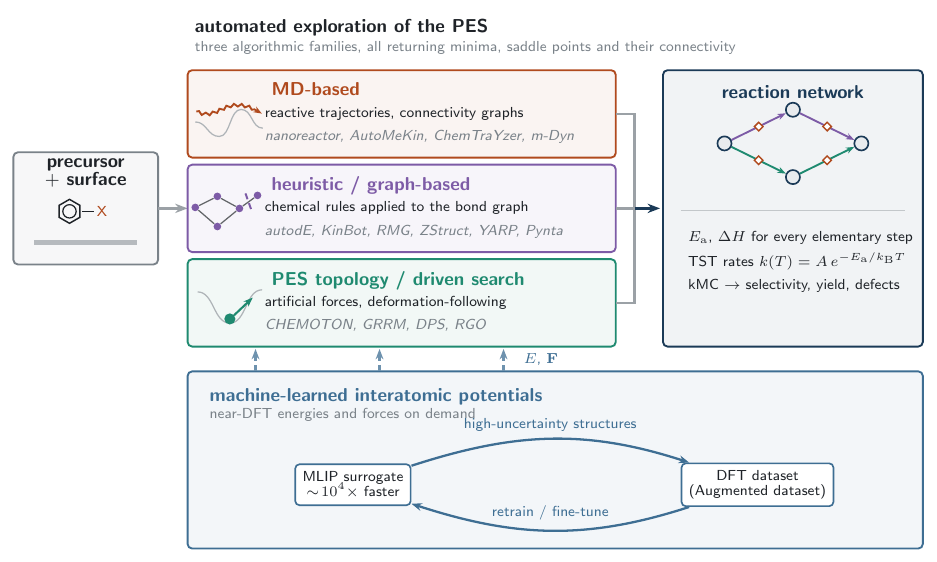}
    \caption{Automated exploration of on-surface reaction networks accelerated by MLIPs. MD-based, heuristic and PES-topology algorithms all return the same reaction network, from which barriers, TST rates and kMC predictions follow. MLIPs supply the energies and forces, with low-confidence structures sent back to DFT in an active-learning loop.}
    \label{fig:mlip_auto}
\end{figure}

	Transitioning from static, user-guided DFT calculations to automated, MLIP-accelerated reaction network exploration can allow atomistic modeling to shift from an explanatory tool to a true predictive pipeline for nanographene manufacturing, as summarised in Fig.~\ref{fig:mlip_auto}.

	\section{Nanographene Self-Assembly on Surfaces: Experiments and Atomistic Modelling}

    While investigations of individual open-shell nanographenes provide fundamental insights into their electronic structure and localized spin states, their integration into ordered molecular assemblies represents an important step toward the realization of functional quantum architectures. In particular, the controlled organization of molecular spin centers on surfaces offers opportunities to establish well-defined intermolecular interactions while potentially enabling individual spin addressing and the development of extended spin networks \cite{Mishra2021Nature,Zhao2024NatNanotechnol}. In such systems, the resulting morphology is governed by the competition between molecule–molecule and molecule–substrate interactions, together with the thermodynamic and kinetic factors controlling molecular diffusion, nucleation, and growth \cite{barth_engineering_2005,Goronzy2018ACSNano}. Molecular organization can, in turn, influence the electronic properties of the assembled system through intermolecular hybridization, substrate-mediated interactions, and collective electronic effects, giving rise to properties that are not necessarily present in the isolated molecular units.
    
    For closed-shell nanographenes, intermolecular organization is largely determined by noncovalent interactions, including dispersion, electrostatic contributions, and substrate-mediated effects, whose relative importance depends on molecular geometry, chemical functionalization, and adsorption conditions. Open-shell nanographenes are subject to the same interactions, but their localized spin states introduce additional electronic degrees of freedom that may become relevant upon aggregation. In particular, intermolecular proximity and relative molecular orientation can modify the overlap between spin-bearing orbitals, giving rise to magnetic exchange interactions whose magnitude and sign depend on the electronic structure and local environment \cite{Krane2023}. Coupling to the supporting substrate can further modify these interactions through electronic screening, charge transfer, and hybridization. Consequently, the relationship between molecular organization and magnetic properties becomes particularly important for open-shell nanographenes, although the extent to which spin-dependent interactions influence the assembly process itself remains to be established.
    
\begin{figure}[ht]
    \centering
    \includegraphics[width=1\linewidth]{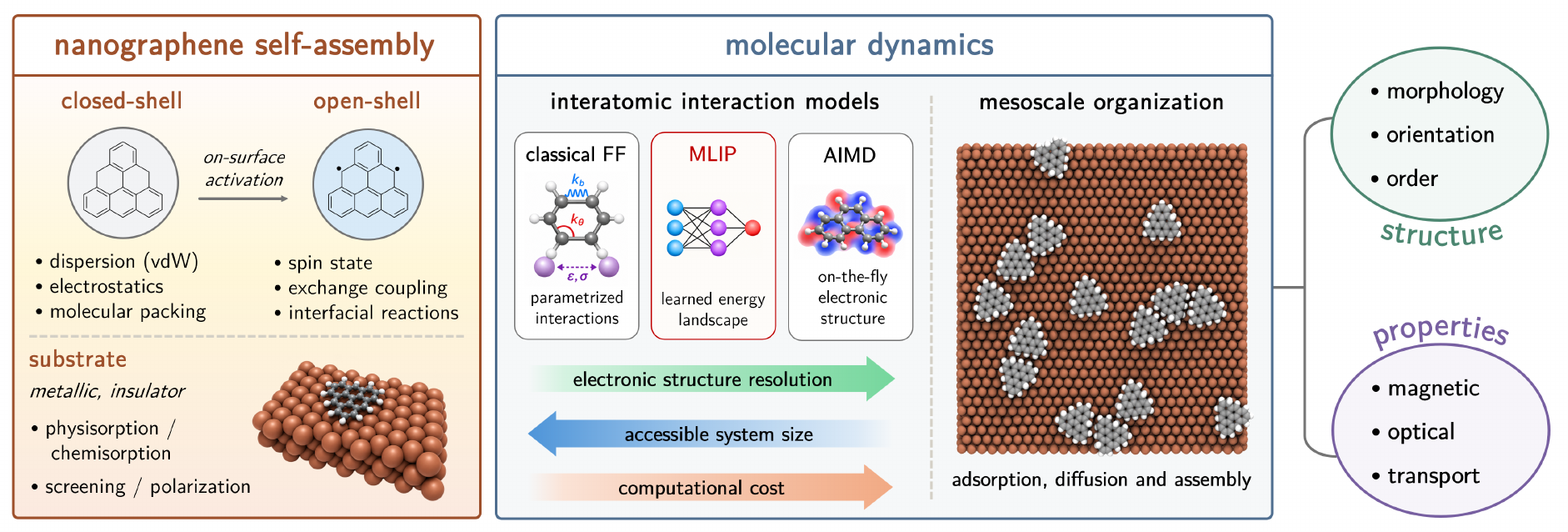}
    \caption{Schematic overview of nanographene self-assembly and its multiscale computational investigation. Molecular electronic configuration, intermolecular interactions, and substrate properties influence adsorption, diffusion, and molecular organization. Classical force fields, machine-learning interatomic potentials, and ab initio MD provide complementary approaches to sampling structural configurations and thermal fluctuations, with different computational costs and levels of electronic-structure description. The resulting morphologies and molecular configurations can serve as inputs for electronic-structure and transport calculations to investigate magnetic, optical, and charge- or spin-transport properties.}
    \label{fig:self_ass}
\end{figure}

    This section first examines experimental strategies for obtaining substrate-supported nanographene assemblies, focusing on the role of surface properties, intermolecular interactions, and molecular packing in determining the resulting structures. Subsequently, computational approaches for investigating the structural evolution of these systems at finite temperature are discussed, with particular emphasis on MD, the limitations of conventional force fields, and the opportunities offered by MLIPs for modeling open-shell nanographenes. Finally, we discuss how structural configurations and trajectories obtained from these simulations can be combined with electronic-structure methods to investigate magnetic exchange networks, optical excitations, and charge- and spin-transport properties, highlighting both established computational strategies and the challenges associated with their application to extended magnetic nanographene assemblies. Figure~\ref{fig:self_ass} summarises this workflow.

	\subsection{Experimental Assembly of Closed- and Open-Shell Nanographenes}
	Experimental advances in the synthesis and characterization of atomically precise nanographenes have enabled increasingly detailed investigations of their adsorption, diffusion, and self-assembly on surfaces. Under ultrahigh-vacuum conditions, molecular deposition and on-surface synthesis on coinage-metal substrates, particularly Au(111), Ag(111), and Cu(111), provide access to individual nanographenes and, under suitable conditions, ordered molecular assemblies \cite{pavlivcek2017synthesis,clair_controlling_2019}. The formation of extended structures depends not only on the intrinsic properties of the molecular building blocks but also on their interactions with the supporting surface, which influences molecular mobility, adsorption geometry, and the stability of competing arrangements. Alternatively, insulating buffer layers, such as h-BN and NaCl, or weakly interacting supports, such as graphene and graphite, can reduce electronic coupling to the underlying substrate, although their suitability for promoting ordered assembly depends on the specific molecule–substrate combination \cite{Joshi2014ACSNano,pavlivcek2017synthesis}.
    
    Surface-confined self-assembly can also play an active role in the fabrication of nanographenes by controlling the spatial organization of molecular precursors before chemical transformation. Depending on molecular coverage and intermolecular interactions, precursor preorganization may favor specific reaction pathways or inhibit competing processes. Consequently, molecular self-assembly and on-surface synthesis should not be regarded as entirely independent phenomena, since the organization of the precursors may influence both the structural selectivity and the final morphology of the resulting nanostructures \cite{clair_controlling_2019,Tenorio2026ChemSci}.
	
	\subsubsection{Role of the Substrate in Molecular Organization}
    
    The substrate provides a spatially varying potential-energy landscape that governs the adsorption, diffusion, nucleation, and growth of molecular assemblies. The balance between molecule–substrate and intermolecular interactions determines whether deposited nanographenes remain isolated, form disordered aggregates, or organize into extended structures. In particular, the strength and corrugation of the adsorption potential influence molecular mobility and the accessibility of different packing configurations. In the physisorption regime, adsorption is predominantly governed by noncovalent interactions, generally preserving molecular electronic states and potentially facilitating reversible rearrangement. Conversely, chemisorption involves significant molecule–substrate hybridization, which can modify molecular geometry, charge distribution, and spin polarization. For instance, [7]-triangulene on Cu(111) exhibits strong molecule–substrate hybridization and quenching of its intrinsic open-shell character \cite{Mishra2021Nanoscale}. Weakly corrugated adsorption landscapes may facilitate molecular rearrangement and the formation of ordered domains, whereas strong site-specific adsorption can restrict diffusion and favor kinetically trapped structures. Nevertheless, strong molecule–substrate interactions may also promote ordering when the surface provides well-defined preferential adsorption sites or directions \cite{barth_engineering_2005,Ruffieux2007JACS}. 
    
    A representative example of substrate-directed organization is provided by the Au(111) herringbone reconstruction, which generates a periodically modulated adsorption landscape containing fcc and hcp regions separated by discommensuration lines and characteristic elbow sites. These structurally inequivalent regions can influence the adsorption and nucleation of polycyclic aromatic molecules. Recently, El Kari\textit{ et al.} investigated the deposition of large C96-based nanographenes on Au(111), observing preferential adsorption at the elbows of the herringbone reconstruction at low coverage. Increasing the molecular coverage promoted the formation of extended striped molecular islands, with orientations influenced by the underlying substrate symmetry \cite{kari_self-assembly_2025}. This example illustrates how substrate reconstruction and intermolecular interactions can cooperate in determining the spatial organization of nanographene assemblies. 
    
    Surface defects and step edges provide additional opportunities for controlling molecular positioning and orientation. Their locally modified coordination environment can stabilize specific adsorption geometries and influence molecular growth directions. Ruffieux \textit{et al.} demonstrated this effect for hexa-peri-hexabenzocoronene (HBC) on Au(111) \cite{Ruffieux2007JACS}, where scanning tunneling microscopy revealed preferential adsorption at monatomic steps with a well-defined molecular orientation. First-principles calculations identified kink sites as energetically favorable adsorption positions, providing an atomistic explanation for the experimentally observed selectivity. The influence of surface defects on molecular packing was also demonstrated by Beyer \textit{et al.}, who showed that HBC films adopt different molecular orientations and packing arrangements on pristine and defective graphite \cite{Beyer2014ACSApplMaterInterfaces}. Beyond naturally occurring defects, single-atom alloys (SAAs), consisting of isolated dopant atoms embedded in a metallic host, provide chemically distinct adsorption sites that can selectively stabilize molecular adsorbates \cite{Chen2025ACSCatal}. This ability suggests a potential strategy for imposing orientational constraints on planar aromatic molecules to potentially reduce the number of competing molecular arrangements during subsequent assembly. 
    
    A different templating mechanism can arise on atomically thin overlayers, such as hexagonal boron nitride (h-BN) and graphene, supported on metallic substrates. Lattice mismatch between the overlayer and the underlying metal can generate Moiré superstructures characterized by spatial variations in adsorption energy and local electrostatic potential. These variations can create preferential adsorption sites and direct the organization of molecular adsorbates. For instance, Joshi \textit{et al.} demonstrated that the electronically modulated landscape of h-BN/Cu(111) can template the self-assembly of porphine and tetracyanoquinodimethane molecules, while simultaneously modifying their energy-level alignment \cite{Joshi2014ACSNano}. More directly relevant to magnetic nanographenes, Catarina \textit{et al.} observed preferential adsorption of phenalenyl dimers along the pore rims of h-BN/Rh(111), while demonstrating the preservation of their magnetic exchange interactions \cite{Catarina2024NanoLett}. Similar opportunities for controlling molecular organization have been demonstrated on graphene/metal Moiré templates \cite{Stradi2014}. Together, these studies establish a relevant strategy for combining spatial organization with reduced electronic coupling to the underlying metal. Its extension to ordered magnetic nanographene arrays, however, requires demonstrating that the templating interactions preserve the desired molecular spin states.
    
    Taken together, these studies show that the substrate influences molecular assembly through several interconnected mechanisms, including preferential adsorption, diffusion barriers and reconstruction-induced confinement. The resulting morphology therefore reflects not only the energetic stability of different packing arrangements but also the kinetic pathways by which these configurations become accessible. Establishing the relative importance of these contributions is essential for predicting the formation of ordered nanographene assemblies under experimentally relevant conditions.
	
	\subsubsection{Closed- vs. Open-shell Assembly}
	Closed-shell nanographenes provide well-established systems for investigating the relationship between molecular structure, intermolecular interactions, and surface-supported organization. Their assembly is generally governed by dispersion, electrostatic and substrate-mediated interactions, together with steric effects associated with molecular shape and peripheral functionalization. For molecules adsorbed recumbently to a surface, lateral packing is primarily determined by interactions between neighboring molecular edges and their adsorption registries, whereas face-to-face $\pi$-stacking becomes particularly relevant in multilayer structures and bulk aggregates \cite{wu_graphenes_2007}.
    
    The coronene family represents a useful platform for examining these effects. In particular, HBC and its derivatives have been extensively investigated as model discotic aromatic systems whose molecular symmetry and peripheral substituents influence their adsorption and packing behavior. Scanning tunneling microscopy studies of extended polycyclic aromatic hydrocarbons on Cu(111) demonstrated that molecule–substrate interactions can largely determine molecular orientation and lattice commensurability, while intermolecular interactions contribute to the packing density and structural organization of the resulting monolayers \cite{ruffieux_self-assembly_2006}. Interestingly, intermolecular interactions are not necessarily attractive. Wagner et al. observed a coverage-dependent lattice parameter for HBC monolayers on Au(111), which they attributed to repulsive interactions arising from a balance between electrostatic contributions, adsorption-induced dipoles, and substrate-screened dispersion forces \cite{Wagner2010PRB}.
    
    Increasing the size of the conjugated framework provides further opportunities to engineer the electronic properties of assembled nanographenes. A particularly illustrative example is circumcoronene, whose on-surface synthesis and self-assembly on Cu(111) were reported by Telychko \textit{et al.} \cite{telychko_ultrahigh-yield_2021}. The molecules form an extended supramolecular superlattice stabilized by interactions with the metallic substrate. Importantly, the resulting periodic electrostatic landscape confines the surface-state electrons of Cu(111), generating an electronic Kagome–honeycomb lattice characterized by emergent flat bands. This example demonstrates that nanographene self-assembly can produce collective electronic phenomena through substrate-mediated confinement, even without invoking magnetic interactions between the molecular building blocks.
    Recent studies on extended nanographenes have demonstrated that substrate-directed self-assembly remains achievable for larger molecular frameworks, despite their increased structural complexity and stronger intermolecular interactions, highlighting the possibility of constructing ordered architectures from increasingly graphene-like building blocks \cite{wu_graphenes_2007,kari_self-assembly_2025}.
    
    The transition from closed-shell to open-shell nanographenes introduces additional complexity because the electronic states responsible for molecular magnetism may be sensitive to intermolecular proximity and adsorption conditions. In particular, the overlap between spin-bearing orbitals can generate magnetic exchange interactions, while substrate-induced screening, charge transfer, and electronic hybridization may modify the molecular spin configuration. Consequently, the preservation of open-shell character within an assembly cannot be inferred solely from the electronic structure of an isolated nanographene and must be established for the relevant substrate \cite{Wang2022}, while also considering the potential influence of molecular arrangement. Furthermore, the reactive nature of some open-shell nanographenes may favor intermolecular covalent bond formation, potentially pairing the unpaired electrons and quenching their magnetic moments. However, whether covalent coupling preserves or suppresses the open-shell character depends on the resulting bonding topology and electronic structure.
    
    Experimentally characterized, noncovalently assembled arrays of open-shell nanographenes with demonstrated intermolecular magnetic coupling remain much less established than assemblies of closed-shell polycyclic aromatic systems. One possible strategy consists of organizing chemically stable, closed-shell precursors before generating the corresponding open-shell nanographenes directly on the surface. Pavliček\textit{ et al.} demonstrated the tip-induced dehydrogenation of dihydrotriangulene precursors to generate individual triangulene molecules \cite{pavlivcek2017synthesis}. Extending this approach to preorganized precursor assemblies could, in principle, allow the formation of magnetic molecular arrays while preserving their initial spatial arrangement, although such structural retention and the stability of the resulting arrays remain to be demonstrated. Much of the progress toward interacting magnetic nanographene architectures has instead relied on the controlled synthesis of covalently connected radical units \cite{Krane2023,Cheng2022,Fu2025NatSynth}. Although these studies do not constitute examples of spontaneous supramolecular assembly, they establish the importance of molecular connectivity and environmental effects in determining magnetic interactions between nanographene units.
    
    The development of ordered open-shell nanographene assemblies therefore presents a distinct challenge: molecular organization must be achieved under conditions that preserve the desired spin states while providing controlled intermolecular coupling. Noncovalent assemblies could offer greater flexibility in tuning intermolecular distances, relative orientations, and consequently magnetic exchange interactions, without modifying the molecular $\pi$-conjugated framework. However, this structural flexibility may also introduce thermal disorder and reduce the stability and reproducibility of magnetic coupling. Depending on the molecular electronic structure, intermolecular interactions may produce either desirable exchange coupling or unwanted electronic hybridization and spin quenching. Understanding this relationship requires identifying the accessible molecular packing configurations and evaluating their electronic and magnetic properties, giving the motivation for the computational approaches discussed in the following section.
	
	\subsection{Molecular Dynamics and Interatomic Potentials}
	The structural evolution and self-assembly of nanographene systems can be investigated through MD simulations, which enable the sampling of molecular configurations and thermal fluctuations at length and time scales beyond those accessible by conventional first-principles methods. Their predictive capability, however, depends critically on the underlying interaction model, particularly its description of intermolecular dispersion and molecule–substrate interactions. While classical force fields enable simulations of large molecular assemblies through predefined interaction parameters, their applicability becomes limited when electronic polarization, charge redistribution, or changes in molecular spin states significantly affect the potential energy surface. 
    
    For surface-supported nanographenes, these limitations are particularly relevant because molecular organization depends on the interplay between intermolecular interactions, substrate screening, and adsorption energetics. Open-shell systems introduce additional challenges whenever structural fluctuations modify spin localization, magnetic coupling, or electronic configurations. MLIPs offer a promising route toward extending atomistic simulations to larger systems and longer timescales, although their accuracy depends on the treatment of long-range interactions and the electronic states represented in the training data. The following subsections discuss these challenges, progressing from molecular assemblies and organic–metal interfaces to emerging machine-learning approaches.

	\subsubsection{Molecular Crystals and Condensed-Phase Assemblies}
    While the preceding section focused on surface-supported nanographenes, established applications of MD to nanographene-related systems have primarily concerned molecular crystals and condensed-phase assemblies. MD simulations provide access to the finite-temperature structural and dynamical properties of molecular crystals and condensed-phase assemblies of $\pi$-conjugated organic materials, including nanographene derivatives. Unlike static geometry optimizations, MD allows the investigation of thermal fluctuations in molecular positions, orientations, and conformations, as well as collective processes such as structural relaxation and phase transitions. These properties are particularly relevant for assemblies stabilized by weak intermolecular interactions, where relatively small changes in molecular packing can influence the stability of competing phases. Classical force fields enable the sampling of these configurations over extended length and time scales, provided that the underlying interactions, particularly dispersion and electrostatics, are described with sufficient accuracy.
    
    Among nanographene-related systems, hexa-peri-hexabenzocoronene (HBC) derivatives provide well-established examples of atomistic MD applications. Previous studies investigated the structural organization and dynamics of alkyl-substituted HBC columnar phases \cite{Andrienko2006JCP}, while others reproduced the transition between crystalline herringbone and liquid-crystalline hexagonal arrangements, highlighting the influence of peripheral substitution on molecular ordering \cite{Marcon2008JCP}. Similarly, Ziogos and Theodorou investigated the thermal and mechanical behavior of HBC-based molecular crystals, including their structural stability and phase transitions \cite{Ziogos2015MolPhys}. Beyond structural characterization, MD trajectories can be combined with electronic-structure calculations to investigate the effects of thermal disorder on electronic properties. For instance, fluctuations in intermolecular orbital overlap influence charge-transfer integrals and charge mobility \cite{troisi_dynamics_2006,kirkpatrick_charge_2007}, while variations in excitonic coupling can affect optical excitation dynamics \cite{Arago2015PRL}. Although these studies concern bulk molecular crystals and liquid-crystalline phases, they provide methodological precedents necessary to extend the investigation of nanographene packing and structural dynamics in surface-supported monolayers.
    
    Despite these advances, the application of MD to open-shell nanographene assemblies remains comparatively unexplored. Importantly, the presence of unpaired electrons does not necessarily require an explicitly spin-dependent force field: when the relevant electronic state remains well defined, structural dynamics can be sampled on a single potential-energy surface, provided that the interaction model accurately reproduces its energetics. However, when competing spin configurations contribute significantly to the structural energetics, their coupling to nuclear motion may require spin-dependent potentials or explicit electronic-structure calculations. Changes in molecular charge, spin localization, or intermolecular bonding introduce further complications. Consequently, the extension of MD to open-shell nanographene assemblies requires interaction models that reproduce the relevant structural energetics and, where necessary, account for spin-dependent interactions. Machine-learning interatomic potentials offer a promising route toward this goal, including architectures that explicitly incorporate magnetic degrees of freedom \cite{Yu2024PRB,kalita_aimnet2nse_2026}, as discussed in Section \ref{sec:os_mlip}.

	\subsubsection{Modelling Organic-Metal Interfaces}
	The organic–metal interface represents one of the main challenges for mesoscale simulations of nanographene assemblies. Conventional fixed-charge force fields, such as CHARMM, OPLS-AA, and AMBER, offer the computational efficiency required for large-scale simulations but generally neglect the electronic response of metallic substrates. In particular, molecular adsorption can induce electronic polarization and charge redistribution at the interface, while the conduction electrons of the metal screen electrostatic interactions between charged or polarizable species. These effects are distinct from net charge transfer and orbital hybridization, which may alter the molecular electronic state. Furthermore, the collective electronic response of the metal modifies dispersion interactions, making an accurate description of van der Waals forces essential for predicting adsorption energies, equilibrium heights, and the balance between molecule–molecule and molecule–substrate interactions \cite{Ruiz2012PRL}.
    
    To overcome some of these limitations, specialized interface force fields have been developed. GolP-CHARMM \cite{wright_golp-charmm_2013}, originally designed for biomolecular adsorption on gold, incorporates metal-specific interactions and an explicit classical representation of gold polarization, including image-charge effects. The Interface Force Field (IFF) \cite{heinz_thermodynamically_2013}, instead, emphasizes thermodynamically consistent parametrization of inorganic–organic interactions, particularly electrostatic and van der Waals contributions, to reproduce adsorption and interfacial properties. These approaches improve the description of interfacial organization while retaining the computational efficiency of classical MD.
    
    Nevertheless, such force fields remain limited by their underlying classical descriptions. Standard fixed-topology parametrizations cannot explicitly describe changes in molecular electronic states, including charge transfer, hybridization, or surface-assisted bond formation. These limitations become particularly relevant for open-shell nanographenes, where substrate interactions may modify spin localization or quench molecular magnetic moments, as discussed in Section \ref{sec:chemi_physi}. Thus, while specialized interface force fields can provide useful structural sampling, their applicability must be assessed against reference adsorption geometries and energetics, particularly when dispersion and electronic substrate effects are significant.
	
	\subsubsection{The Open-Shell Challenge and the Machine-Learning Potentials}\label{sec:os_mlip}
    
    The simulation of open-shell nanographene assemblies remains a major challenge for predictive mesoscale modelling. As discussed in the previous subsections, conventional fixed-charge force fields can provide structural sampling but generally cannot describe changes in molecular charge or spin configurations. At the opposite end of the accuracy–efficiency spectrum, ab initio molecular dynamics (AIMD) can account for electronic-structure changes within the chosen theoretical approximation, but its computational cost limits the system sizes and timescales accessible for molecular diffusion and self-assembly. MLIPs offer a promising alternative by approximating first-principles potential-energy surfaces at substantially reduced computational cost. However, their suitability for open-shell nanographenes depends not only on the accuracy of the learned local interactions, but also on how dispersion, long-range electrostatics, and electronic or magnetic degrees of freedom are represented.
    
   \paragraph{Local interatomic potentials} MACE (Many-body Atomic Cluster Expansion) \cite{batatia_mace_2022}  combines equivariant message passing with higher-order many-body representations to model complex atomic environments. SNAP (Spectral Neighbor Analysis Potential) \cite{thompson_spectral_2015} instead employs rotationally invariant bispectrum descriptors to construct interatomic potentials and has been applied extensively to materials systems. In their standard forms, both approaches describe atomic environments within finite interaction cutoffs. For MACE, a radial cutoff of approximately 5 Å is commonly employed, although the value is model-dependent and successive message-passing layers extend the effective receptive field. Such architectures can reproduce short-range chemical interactions and structural rearrangements represented in their training data, but do not automatically capture interactions beyond their effective range. For surface-supported nanographenes, this limitation is particularly important for noncovalent intermolecular interactions and molecule–substrate binding. 
    
    \paragraph{Treatment of dispersion} Long-range dispersion is essential for predicting the relative stability of molecular packing arrangements and adsorption geometries. One strategy is to supplement a local MLIP with an explicit dispersion correction, such as Grimme's D3 method \cite{grimme2010consistent}, which can be combined with an MLIP through the \texttt{pair\_style} hybrid/overlay framework in LAMMPS \cite{Thompson2022CPC}.  However, care must be taken to avoid double counting, since an MLIP trained on dispersion-inclusive DFT energies may already reproduce part of these interactions. The Tkatchenko–Scheffler (TS) approach provides an alternative based on environment-dependent dispersion coefficients derived from the electron density \cite{Tkatchenko2009PRL}, but its application alongside an MLIP requires a compatible procedure for obtaining those coefficients. As mentioned in Section\ref{sec:chemi_physi}, more sophisticated treatments, including many-body dispersion (MBD) methods and nonlocal van der Waals density functionals, can improve the description of collective dispersion effects. This is particularly relevant for adsorption on metallic substrates, where conventional pairwise corrections such as D3 do not explicitly account for collective substrate screening. The vdW-surf method addresses this limitation by incorporating substrate screening into the dispersion treatment and provides an important reference for evaluating adsorption energetics \cite{Ruiz2012PRL}. However, vdW-surf is a density-functional correction scheme rather than a directly interchangeable classical pair potential.
    
    \paragraph{Long-range electrostatics and charge redistribution} Several MLIP architectures explicitly extend local atomic descriptions with nonlocal interactions. Fourth-generation high-dimensional neural-network potentials (4G-HDNNPs) combine environment-dependent atomic electronegativities with charge equilibration, enabling the treatment of nonlocal charge redistribution \cite{Ko2021NatCommun}. Deep Potential Long Range (DPLR) incorporates long-range electrostatic interactions through a representation involving ionic and electronic charge centres \cite{Zhang2022JCP}. These methods address important limitations of strictly local potentials, although neither architecture is inherently designed to represent competing molecular spin states. More recently, MACE-POLAR has extended the MACE framework through explicit long-range electrostatics, polarization, and charge- and spin-dependent response, providing a potentially relevant direction for molecular assemblies in which electrostatic induction and electronic-state changes are significant \cite{batatia_mace-polar-1_2026}.
    
    \paragraph{Open-shell electronic states and magnetic interactions} AIMNet2-NSE introduces neural spin-charge equilibration to describe molecular configurations with different charges and spin multiplicities, making it relevant to the modelling of open-shell nanographenes \cite{kalita_aimnet2nse_2026}. A complementary approach is represented by SpinGNN \cite{Yu2024PRB}, which incorporates magnetic degrees of freedom to model coupled spin–lattice interactions in magnetic materials. These methods address different aspects of open-shell behaviour: AIMNet2-NSE focuses on spin-dependent molecular electronic energetics, whereas SpinGNN explicitly represents magnetic configurations and their coupling to structural degrees of freedom. Nevertheless, their applicability to extended nanographene assemblies on metallic substrates requires further assessment.
    
    No single approach discussed here has been established as a transferable solution for the simultaneous treatment of molecular self-assembly, substrate-screened dispersion, charge transfer, and magnetic interactions in open-shell nanographenes. Progress will therefore require not only suitable model architectures and targeted training data, but also validation against adsorption energies and geometries, diffusion barriers, competing packing arrangements, and, where relevant, the relative energies of different spin configurations.

	\subsection{From Simulated Morphologies to Magnetic, Optical, and Transport Properties}
	
    The primary objective of mesoscale simulations of nanographene self-assembly is to investigate the formation and structural organization of molecular assemblies, including their preferred packing arrangements and structural disorder. Beyond these structural predictions, MD simulations provide access to finite-temperature fluctuations and distributions of molecular configurations, provided that the interaction model is sufficiently accurate and the relevant configurations are adequately sampled. However, reproducing experimentally observed morphologies also requires accounting for preparation conditions and kinetic effects, since deposition, thermal annealing, and subsequent cooling may lead to metastable or kinetically trapped arrangements that are not represented by equilibrium sampling alone. The resulting simulated morphologies and molecular configurations can then serve as structural inputs for electronic-structure and transport calculations, enabling the prediction of magnetic, optical, and electronic properties and, crucially, their dependence on molecular organization, thermal fluctuations, and disorder. This multiscale strategy offers a route to establishing quantitative relationships between self-assembly and the emergent properties of nanographene-based quantum architectures.
    
    \paragraph{Magnetic coupling networks} For open-shell nanographene assemblies, one of the key targets is the characterization of intermolecular magnetic interactions within the assembled structure. By extracting representative molecular configurations from MD trajectories, spin-polarized DFT and broken-symmetry approaches can be employed to estimate exchange coupling constants (J) between neighboring spin centers. The sensitivity of these interactions to molecular geometry has already been demonstrated in triangulene-based dimers, where variations in the relative orientation of the radical units can substantially modify magnetic coupling \cite{Yu2023}, although this effect remains to be quantified for noncovalently assembled nanographene systems. Mapping these values across a molecular array would allow the construction of configuration-dependent exchange networks, providing insight into the sensitivity of magnetic interactions to thermal disorder and molecular packing. Assessing collective magnetic ordering and its thermal stability would additionally require an appropriate spin Hamiltonian and statistical-mechanical treatment.
    
    \paragraph{Optical and spectroscopic properties} The structural fluctuations captured by MD trajectories can also serve as input for predicting optical signatures of nanographene assemblies. Excited-state calculations, such as time-dependent DFT, can be performed on representative configurations to evaluate absorption spectra, excitonic states, and the contribution of structural disorder to spectral broadening, although a complete description of temperature-dependent spectra may additionally require vibronic and dynamical effects. The importance of this approach has been demonstrated in organic molecular crystals, where thermal motion produces substantial fluctuations in excitonic coupling \cite{Arago2015PRL}. For nanographenes, analogous calculations could investigate how molecular conformations and intermolecular arrangements influence excitation energies, excitonic interactions, and optical transitions involving different spin states, although appropriate excited-state methods are required when significant multireference character is present.
    
    \paragraph{Charge and spin transport simulations} The electronic couplings extracted from MD-sampled structures can be used as input for charge- and spin-transport models. For closed-shell nanographenes and related organic semiconductors, fluctuations in intermolecular transfer integrals can be incorporated into hopping or transient-localization models to evaluate charge mobility \cite{troisi_dynamics_2006,kirkpatrick_charge_2007,Fratini2017NatMater}. For open-shell assemblies, an analogous multiscale strategy could combine spin-dependent electronic-structure calculations on representative MD configurations with suitable transport models, selected according to the relevant transport regime. Nonequilibrium Green's function (NEGF) methods can describe coherent or partially coherent spin-dependent transport \cite{rocha_spin_2006}, whereas hopping approaches are more appropriate when charge carriers are localized and transport proceeds through thermally activated transitions between molecular sites \cite{bobbert_theory_2009,Tolkachev2025PRB}. Such simulations could clarify how molecular packing, structural disorder, and intermolecular magnetic interactions affect spin-dependent transport, although their quantitative applicability to extended nanographene networks remains to be established.
	
	\section{Conclusions}
	The rational design and scalable fabrication of nanographene-based quantum materials represent a convergence of surface chemistry, electronic structure theory, and mesoscale simulations to access thermodynamical parameters. Over the past decade, the field has evolved from observing isolated, gas-phase-like molecules on inert surfaces to actively engineering spin states, reaction pathways, and supramolecular topologies at organic–metal interfaces.
	
	Achieving predictive control over these complex architectures requires bridging the divide between electronic-scale accuracy and mesoscale spatiotemporal dynamics. Looking forward, several critical bottlenecks and emerging opportunities will define the computational and experimental landscape:
	
	\begin{enumerate}
		\item \textit{Unified Spin-Active Machine Learning Potentials}: While MLIPs have revolutionized total-energy mapping for materials, a persistent challenge is the simultaneous modeling of variable charge transfer, polarization, and non-collinear spin degrees of freedom. Frameworks that integrate explicit spin-charge equilibration (e.g., AIMNet2-NSE) with long-range polarizability and equivariant message passing (e.g., MACE-POLAR) must be further developed and benchmarked against extended metal–organic interfaces.
		
		\item \textit{High-Throughput Kinetic Screening vs. Experimental Validation}: Automated reaction path discovery algorithms coupled with MLIPs are now capable of generating extensive kinetic networks for complex polycyclic precursors. The key bottleneck is translating these theoretical networks into actionable experimental blueprints under non-equilibrium synthesis conditions, requiring tighter integration with fast operational spectroscopic techniques (e.g., time-resolved XPS, LT-STM/AFM).
		
		\item \textit{Preserving Quantum Coherence in Extended Assemblies}: A central goal for quantum information applications is building extended $J$-coupling arrays from open-shell monomers without quenching their localized spin states. Mesoscale MD trajectories coupled with non-equilibrium transport and broken-symmetry DFT offer a clear path to evaluate how dynamic thermal disorder suppresses or stabilizes collective magnetic phases.
	\end{enumerate}
	
	Ultimately, transitioning from an explanatory modeling paradigm to a predictive, blueprint-to-fabrication workflow will rely on seamlessly combining automated potential energy surface exploration, spin-aware neural network potentials, and finite-temperature transport models. Addressing these open challenges will accelerate the integration of atomically precise nanographenes into scalable, spin-based quantum technologies.

\backmatter

\section*{Declarations}

\bmhead{Funding}
The authors acknowledge ``Italian
Science Fund -- FIS~2 Call'' through the iPAWNS starting grant
(CUP: B53C24009560001)

\bmhead{Competing interests}
The authors declare no competing interests.

\bmhead{AI declaration}
During the preparation of this manuscript, the authors used generative AI
tools to assist with language editing, reference formatting and checking,
and \LaTeX{} preparation. No AI tool was used to generate data,
results, or scientific conclusions. All AI-assisted output was reviewed,
verified, and edited by the authors, who take full responsibility for the
content of this article.

\bmhead{Author contributions}
All authors contributed equally to the conception, literature survey, and
writing of this review. All authors read and approved the final manuscript.

\bibliography{library}

\end{document}